\documentclass[twopage]{elsart}

\begin{document}

\input epsf

\begin{frontmatter}
\title{Thermodynamics of the two--dimensional O(3)
$\sigma$--model with fixed--point lattice action}

\author[Zurich]{S.~Spiegel}
\address[Zurich]{Institut f\"ur Theoretische Physik, Universit\"at Z\"urich,
Winterthurerstr. 190, CH-8057 Z\"urich, Switzerland}

\begin{abstract}
The free energy density of the two--dimensional O(3) non--linear 
$\sigma$--model is calculated at finite temperature and finite spatial
extent. We make both an analytic calculation in the perturbative regime and a
Monte--Carlo study at low temperatures. We show that using the
fixed--point action instead of the standard Wilson action leads to a great
reduction of the cut--off effects. 
\end{abstract}

\end{frontmatter}

\newpage

\section{Introduction}  \label{intro}
The calculation of thermodynamical quantities in Yang--Mills gauge
theories using lattice regularization
suffers from huge cut--off effects. Decreasing the lattice spacing $a$,
 the cost of such calculations grows  with
 $\left[ \frac{1}{a} \right]^{\epsilon}$, where in SU(3) gauge
theory $\epsilon$ is around 10. This requires a very careful
analysis  to remove the cut--off effects. 
 Therefore in
field theory one looks for improved actions which give good results 
even on lattices with  bad resolution.
 Wilson's idea of the renormalization group
(RG) \cite{Kogut, Wilson}  aims at that direction
: one integrates out the short--distance degrees
of freedom and replaces them by a smaller set of effective degrees of freedom,
which describe long--distance physics.
One
possibility is to use a blocking procedure introduced originally by Kadanoff.
Theoretically there exists a perfect lattice action,
that predictions are absolutely free
of any cut--off effects. We use here the fixed--point (FP)
action, which is the classical perfect action and
is expected to be a good approximation for the
perfect action. For the O(3) non--linear 
$\sigma$--model, the model we are considering here, the FP action
was developed in \cite{Hase1}. Recently in a nice
 work \cite{cut}, Alessandro Papa pointed out
 the great advantage of the FP action for SU(3) gauge theory:
The free energy density near the phase transition point showed an
impressive reduction of cut--off effects compared to earlier
results derived by other lattice actions \cite{Boyd}.
However the few points calculated
in that paper do not allow  a complete
 view of  the behaviour in the decrease
of cut--off effects. In the case of the $\sigma$--model we can
continue this project and give the free energy density over a large
temperature region. Hence
we check the
cut--off effects arising from different lattice actions both in
the high temperature regime (perturbation theory) and at
low temperatures using Monte--Carlo (MC) simulations.

The Euclidean continuum version of the partition function for the O(3) 
non linear  $\sigma$--model in two dimensions reads as follows:
\begin{equation}
Z=\int\! \! {\mathrm D} {\mathbf \mathit S}(x) \; 
{\mathrm e}^{ -\beta {\mathcal A}({\mathbf \mathit  S})}
\label{eq:Z}
\end{equation}
where ${\mathbf \mathit S}$ is a three component vector with 
${\mathbf \mathit S}^{2}(x)=1$ and measure
\begin{equation}
{\mathrm D} {\mathbf \mathit S}(x)=
\prod_x {\mathrm d}^{3} {\mathbf \mathit S}(x) \;
\delta( {\mathbf \mathit S}^{2}(x) - 1) \quad .
\label{eq:meas}
\end{equation}
The term $\beta {\mathcal A}({\mathbf \mathit S})$ is the continuum
action with periodic boundary conditions for the field 
${\mathbf \mathit S}$
at finite temperature $T=1/L_{\mathrm t}$ and spatial extent $L_{\mathrm s}$.
\begin{equation}
\beta {\mathcal A}_{\mathrm {cont}}({\mathbf \mathit S})=
\frac{\beta}{2} \int_{0}^{L_{\mathrm t}} \! \! {\mathrm d} x_{0}
\! \! \int_{0}^{L_{\mathrm s}} \! \! {\mathrm d} x_{1} 
\; \partial_{\mu} {\mathbf \mathit S}(x) \partial_{\mu} 
{\mathbf \mathit S}(x)
\label{eq:cact}
\end{equation}
The free energy density ${\mathbf f}$ is given by the relation
\begin{equation}
\label{eq:free}
Z= {\mathrm e}^{ -{\mathbf f} V}
\end{equation}
where $V$ is $L_{\mathrm t} \times L_{\mathrm s}$
and  ${\mathbf f}$ depends on both $L_{\mathrm t}$ and
$L_{\mathrm s}$. To get  the
temperature and spatial size
 dependent, physical part of this quantity we subtract the
contribution resulting from the square box, in which we set 
$L_{\mathrm t}$ equal to $L_{\mathrm s}$. In the perturbative
 regime
 and for infinite spatial extent we get
$-{\mathbf f}/T^{2}=\pi /3$ which is the two--dimensional case of an
ideal gas of massless scalar particles.

This letter consists of the following parts: in section \ref{fix} we give a
short review of the ideas which lead to the FP action in the
$\sigma$--model. The next section presents the perturbative
calculation of the free energy density and a
comparison between different lattice actions.  Section \ref{monte}
 treats the cluster--MC analysis necessary for the
calculation in the non--perturbative, low temperature regime
and shows the corresponding results.

\section{FP action} \label{fix}

We summarize now some important results about the FP action
of the O(3) $\sigma$--model which we use in this letter. For a
detailed discussion we refer the reader to ref. \cite{Hase1}.

We divide our original lattice into $2 \times 2$ blocks labeled by 
$n_{\mathrm B}$. The four spins ${\mathbf \mathit S}_{n}$ inside
a block form,
by taking some mean,  a new spin variable
${\mathbf \mathit R}_{n_{\mathrm B}}$ on a lattice with a lattice unit
 twice as large as
the original one. The new action ${\mathcal A}'({\mathbf \mathit R})$
is defined by integration over the original lattice:
\begin{equation} \label{eq:blocking}
{\mathrm e}^{ -\beta' {\mathcal A}'({\mathbf \mathit R})}
= \int\! \! {\mathrm D} {\mathbf \mathit S}(x) \; 
{\mathrm e}^{ -\beta \left[ \rule[-.5ex]{0ex}{2ex}
 {\mathcal A}({\mathbf \mathit  S})
+ {\mathcal T}({\mathbf \mathit R},{\mathbf \mathit S}) \right]}
\end{equation}
where ${\mathcal T}$ is the kernel of the RG transformation with the
normalization that ensures the invariance of the partition function
\begin{equation} \label{eq:norm}
\int\! \! {\mathrm D} {\mathbf \mathit R}(x) \; 
{\mathrm e}^{ -\beta {\mathcal T}({\mathbf \mathit R},
{\mathbf \mathit S})}=1 \quad .
\end{equation}
In the limit $\beta \rightarrow \infty$ the blocking kernel 
${\mathcal T}$ has the form 
\begin{equation} \label{eq:kernel}
{\mathcal T}({\mathbf \mathit R},{\mathbf \mathit S})=
\kappa \sum_{n_{\mathrm B}}
 \left(
 \left|
\rule[-.5ex]{0ex}{2ex} 
\sum_{n \in n_{\mathrm B}}{\mathbf \mathit S}_{n} 
\right|
- {\mathbf \mathit R}_{n_{\mathrm B}} \cdot
\sum_{n \in n_{\mathrm B}}{\mathbf \mathit S}_{n}
\right) 
\end{equation} 
where $\kappa$ (fixed here to $2$) can be tuned to find a 
short--ranged FP--action. In the limit mentioned above
 equation (\ref{eq:blocking}) reduces to
a saddle point problem, giving
\begin{equation} \label{eq:saddle}
{\mathcal A}'({\mathbf \mathit R}) =
\min_{\{  {\mathit S}\}  }
\left\{
\rule[-.7ex]{0ex}{3.4ex}
{\mathcal A}({\mathbf \mathit S}) + {\mathcal T}
({\mathbf \mathit R},{\mathbf \mathit S})
\right\}  ,
\end{equation} 
and for the FP action
\begin{equation} \label{eq:fix}
{\mathcal A}_{\mathrm {FP}}({\mathbf \mathit R}) =
\min_{\{ {\mathit S} \} }
\left\{
\rule[-.7ex]{0ex}{3.4ex}
{\mathcal A}_{\mathrm {FP}}({\mathbf \mathit S}) + {\mathcal T}
({\mathbf \mathit R},{\mathbf \mathit S})
\right\}  .
\end{equation}
For parametrizing it,  one can write in general
\begin{eqnarray} \label{eq:par}
{\mathcal A}_{\mathrm {FP}}({\mathbf \mathit S}) & = &
- \half \sum_{n,r} \rho (r)(1-{\mathbf \mathit S}_{n}
{\mathbf \mathit S}_{n+r})  \nonumber  \\
& & + \sum_{n_{1},n_{2},n_{3},n_{4}}
c(n_{1},n_{2},n_{3},n_{4})(1-{\mathbf \mathit S}_{n_{1}}
{\mathbf \mathit S}_{n_{2}})
(1-{\mathbf \mathit S}_{n_{3}}{\mathbf \mathit S}_{n_{4}}) +
\ldots   ,
\end{eqnarray}
where it is a significant help, that
the first two
 functions $\rho$ and $c$ can be calculated analytically. The
Fourier transform of $\rho$ has the form
\begin{equation} \label{eq:ro}
\frac{1}{\rho(q)} =
\sum_{l=-\infty}^{+\infty} \frac{1}{(q+2 \pi l)^{2}}
\prod_{i=0}^{1} \frac{\sin^{2}(q_{i}/2)}{(q_{i}/2 + \pi l_{i})^{2}}
+ \frac{1}{3 \kappa} \quad .
\end{equation}
 The summation is over the integer vector $l=(l_{0},l_{1})$
and $(q+2 \pi l)^{2}=(q_{0} + 2 \pi l_{0})^{2} +
(q_{1} + 2 \pi l_{1})^{2}$. Furthermore one can show that $\rho$
fulfills the FP equation for a free scalar theory with a
gaussian block--transformation.

Beside the two functions $\rho$ and $c$ the resulting couplings 
were found in \cite{Hase1} by a numerical fitting procedure and
recently improved in \cite{Ruedi} to control the topological effects.
We use the  parametrization obtained in \cite{Ruedi}. It consists of 
 24 different couplings which  can be put on  a
$2 \times 2$ lattice. When working with such an action to measure the
free energy density one expects to see cut--off effects,
due to the following three points:
\begin{itemize}
\item The FP action is not the perfect action.
\item In the simulation a parametrized form of the FP
action is used which introduces some parametrization errors.
\item Finally, on extremely small lattices ($N_{\mathrm t}=2$) the
action is disturbed by the small size, which causes cut--off
effects that  decrease exponentially with $N_{\mathrm t}$.
\end{itemize}

\section{Perturbative results} \label{presult}

In a normal perturbative treatment 
 of eq. (\ref{eq:Z}), one introduces
a Faddeev--Popov term to avoid the zero mode problem \cite{Leutwyler}.
One arrives finally, when changing to Fourier space, at the following
form of the partition function:
\begin{equation}
{\mathrm e}^{-\sum_{k}'\;
\ln k^{2} + \; \ln V } 
\label{eq:zper}
\end{equation}
where the sum here just leaves away the unwanted zero mode
$k=(0,0)$, $\sum_{k}' = \sum_{k,k\neq(0,0)}$. As  a remainder
of the Faddeev--Popov term we have to keep the logarithm of the
volume in the partition function. Due to equation (\ref{eq:cact})
the Fourier sum above is discreet and one can use any regularization
scheme to extract the continuum value ${\mathbf f}_{\mathrm {cont}}$.
 When using 
lattice regularization  one must replace the continuum 
propagator $k^{2}$ in eq. (\ref{eq:zper}) with the corresponding
lattice propagator, for example with the function $\rho$ of
eq. (\ref{eq:ro}) in the FP case. We then check the 
cut--off effects by considering the ratios
${\mathbf f}_{\mathrm W}/{\mathbf f}_{\mathrm {cont}}$,
${\mathbf f}_{\mathrm S}/{\mathbf f}_{\mathrm {cont}}$,
${\mathbf f}_{\mathrm {fi}}/{\mathbf f}_{\mathrm {cont}}$ and
${\mathbf f}_{\mathrm {FP}}/{\mathbf f}_{\mathrm {cont}}$, where the 
abbreviations ''W, S `` denote the use of
Wilson and Symanzik  action,
 whereas ''${\mathrm {fi}}$``
means an action with a nearest neighbour
and a diagonal coupling, whose ratio is the same as in the FP
action. Although there is no problem to give ${\mathbf f}$ as a
function of $\zeta$, $\zeta=L \cdot T=L_{\mathrm s}/L_{\mathrm t}$, 
we present the cut--off effects for $\zeta=3$.
 Table \ref{tab:num}  gives the results  ${\mathbf f}_{\mathrm {cont}}
/ {\mathbf f}_{\mathrm {latt}}$ for different lattice schemes, a  different
number of points $N_{\mathrm t}$ in the time direction
(i.e. for different
resolutions) and $N_{\mathrm s}=3 \times N_{\mathrm t}$.
An approximation of the decrease of the cut--off effects can then be
given through a function of the form:
\begin{equation}
1+a\cdot\frac{1}{N_{\mathrm t}^{2}} +
b\cdot\frac{1}{N_{\mathrm t}^{4}} +
c\cdot\frac{1}{N_{\mathrm t}^{6}} + \ldots \quad .
\end{equation}
The Wilson action 
shows the expected quadratic decrease (fig. 2a), whereas in the
Symanzik case (fig. 2b) the fit is consistent with $a=0$.
Figure 2c shows the behaviour of ${\mathbf f}_{\mathrm {fi}}$ and
figure 2d finally presents a perfect exponential decrease of the 
cut--off effects for the FP action. The smooth curve is
given by $1+A\cdot\exp(-BN_{\mathrm t})$, where $A$ and $B$ are
constants.

\begin{table}[h]
\begin{center}
\caption{Numerical values for different lattice actions with
$N_{\mathrm s}=3 \times N_{\mathrm t}$}.
\label{tab:num}
\vspace*{6ex}
\begin{tabular}{ccr@{.}lr@{.}lr@{.}lr@{.}l}
\multicolumn{2}{c}{\# $N_{\mathrm t}$} &
\multicolumn{2}{c}{ ${\mathbf f}_{\mathrm W}/{\mathbf f}_{\mathrm {cont}}$} &
\multicolumn{2}{c}{${\mathbf f}_{\mathrm {cont}}/{\mathbf f}_{\mathrm S}$} &
\multicolumn{2}{c}{${\mathbf f}_{\mathrm {cont}}/{\mathbf f}_{\mathrm {fi}}$} &
\multicolumn{2}{c}{${\mathbf f}_{\mathrm {cont}}/{\mathbf f}_{\mathrm{FP}}$} \\
2 && 1&1606 & 0&964058 & 1&016701 & 1&006314 \\
4 && 1&0455 & 1&001507 & 1&004622 & 1&000016 \\
6 && 1&0182 & 1&000155 & 1&001773 &  1 &000000      \\
8 && 1&0098 & 1&000024 & 1&000729 &  1 &000000      \\
10 && 1&0062 & 1&000007 & 1&000240 &1 &000000       \\
\end{tabular}
\vspace*{6ex}
\end{center}
\end{table}

\begin{figure}[h]
\begin{center}
\leavevmode
\epsfxsize=\textwidth
\epsfbox{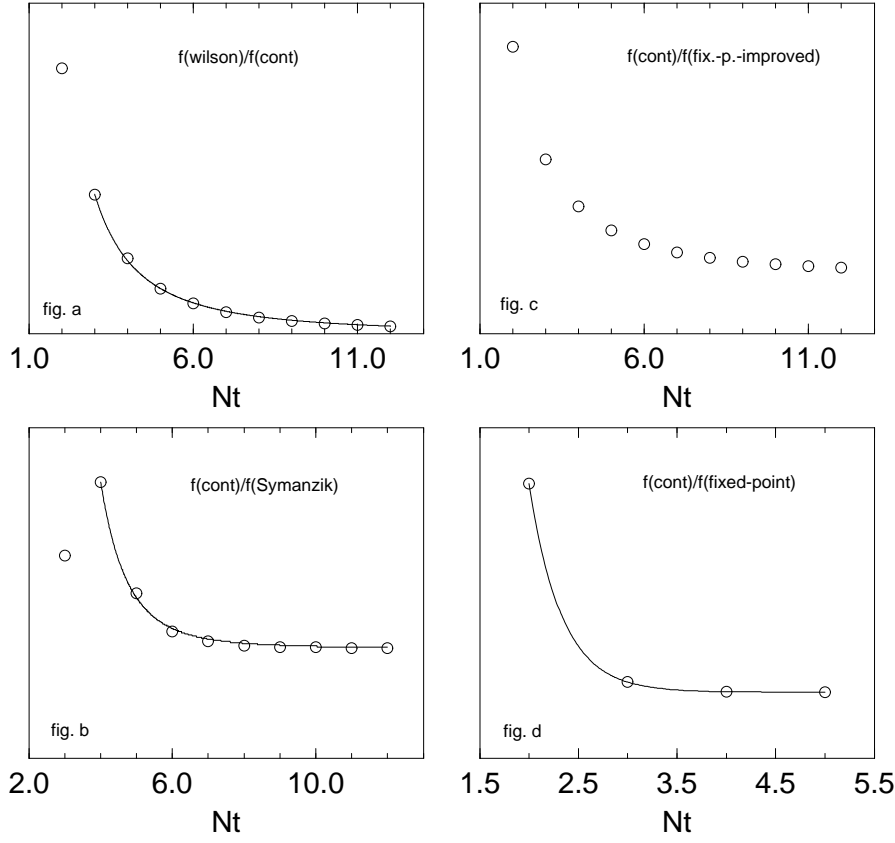}
\caption{Figures $a$--$d$ show the disappearence of the
cut--off effects for different lattice actions. The smooth curves are
fitted functions explained in the text.}
\label{fig:decrease}
\end{center}
\vspace*{6ex}
\end{figure}

\section{MC methods}
\label{monte}

In a MC simulation the partition function cannot be measured
directly, on the other hand, using equations
 (\ref{eq:Z},\ref{eq:free}),  the derivative of the logarithm
of $Z$ with respect to $\beta$ gives:
\begin{equation}
\frac{\partial}{\partial \beta} \; \ln Z =
-Z^{-1} \!  \int \! \! {\mathrm D} {\mathbf \mathit S} \;
{\mathcal A}({\mathbf \mathit S}) \;
{\mathrm e}^{-\beta {\mathcal A}({\mathbf \mathit S})} \quad .
\label{eq:dbeta}
\end{equation}
One obtains
\begin{equation}
-\frac{{\mathbf f}}{T^{2}} =
\int_{\beta_{0}}^{\beta} \! \! {\mathrm d} \beta' \;
\left[
\frac{\langle {\mathcal A}_{\mathrm {vac}} \rangle}{\zeta^{2}} -
\frac{\langle {\mathcal A}_{\mathrm t} \rangle}{\zeta}
\right]  
\label{eq:thermo}
\end{equation}
where $\langle {\mathcal A}_{\mathrm t} \rangle$ denotes the
expectation value
of the action on the lattice $N_{\mathrm t} \times N_{\mathrm s}$ and
$\langle {\mathcal A}_{\mathrm {vac}} \rangle$ the same on the
square lattice which we subtract as vacuum contribution. 

\begin{figure}
\begin{center}
\leavevmode
\epsfxsize=0.8\textwidth
\epsfbox{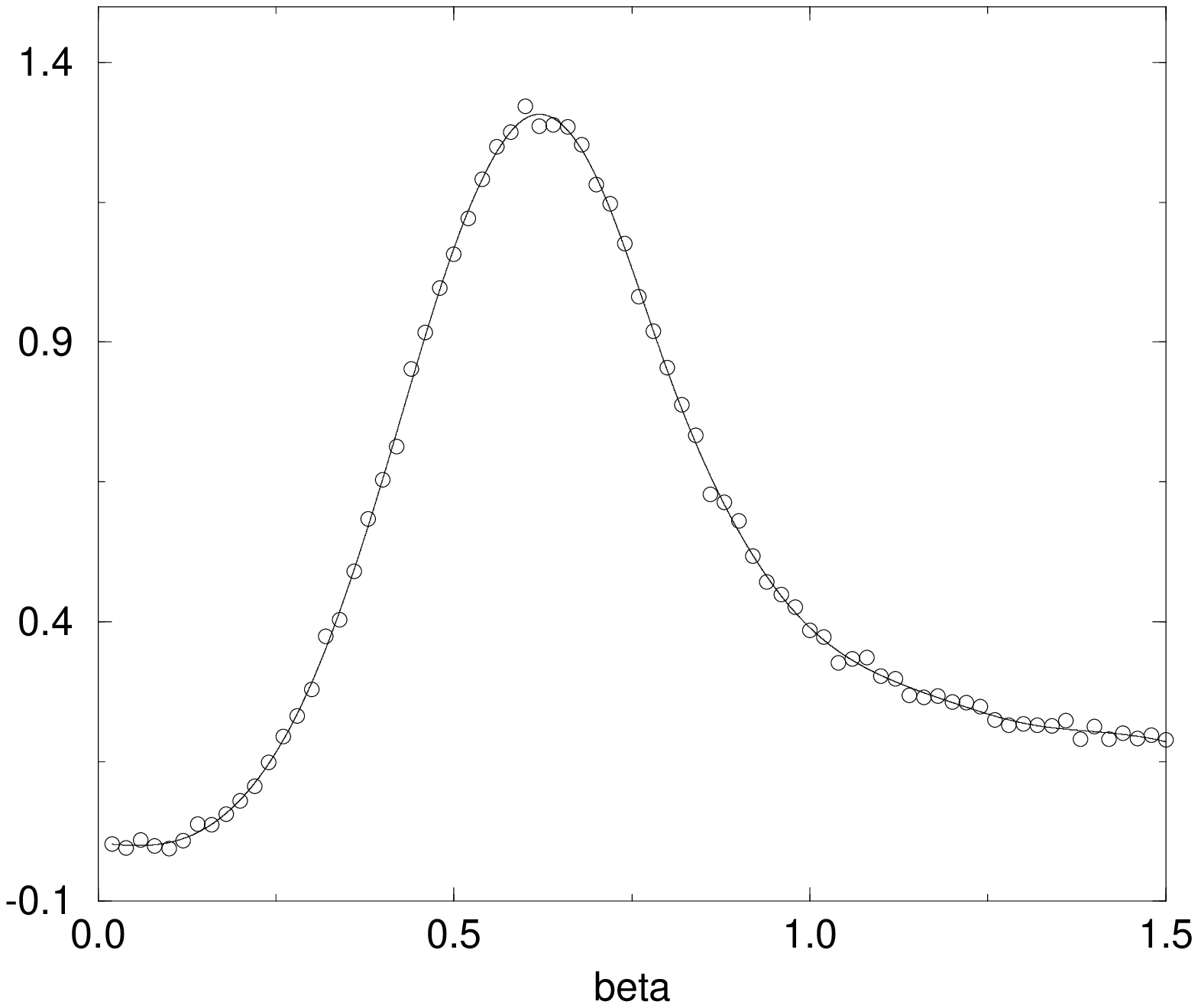}
\caption{The signal of the integrand in eq. (\ref{eq:thermo})
for a $4 \times 12$ lattice  is 
plotted against the values of $\beta$. 
 The smooth curve is the spline interpolation we used for
the integration.}
\label{fig:deltas}
\end{center}
\vspace*{6ex}
\end{figure}

\begin{figure}[h]
\begin{center}
\leavevmode
\epsfxsize=0.8\textwidth
\epsfbox{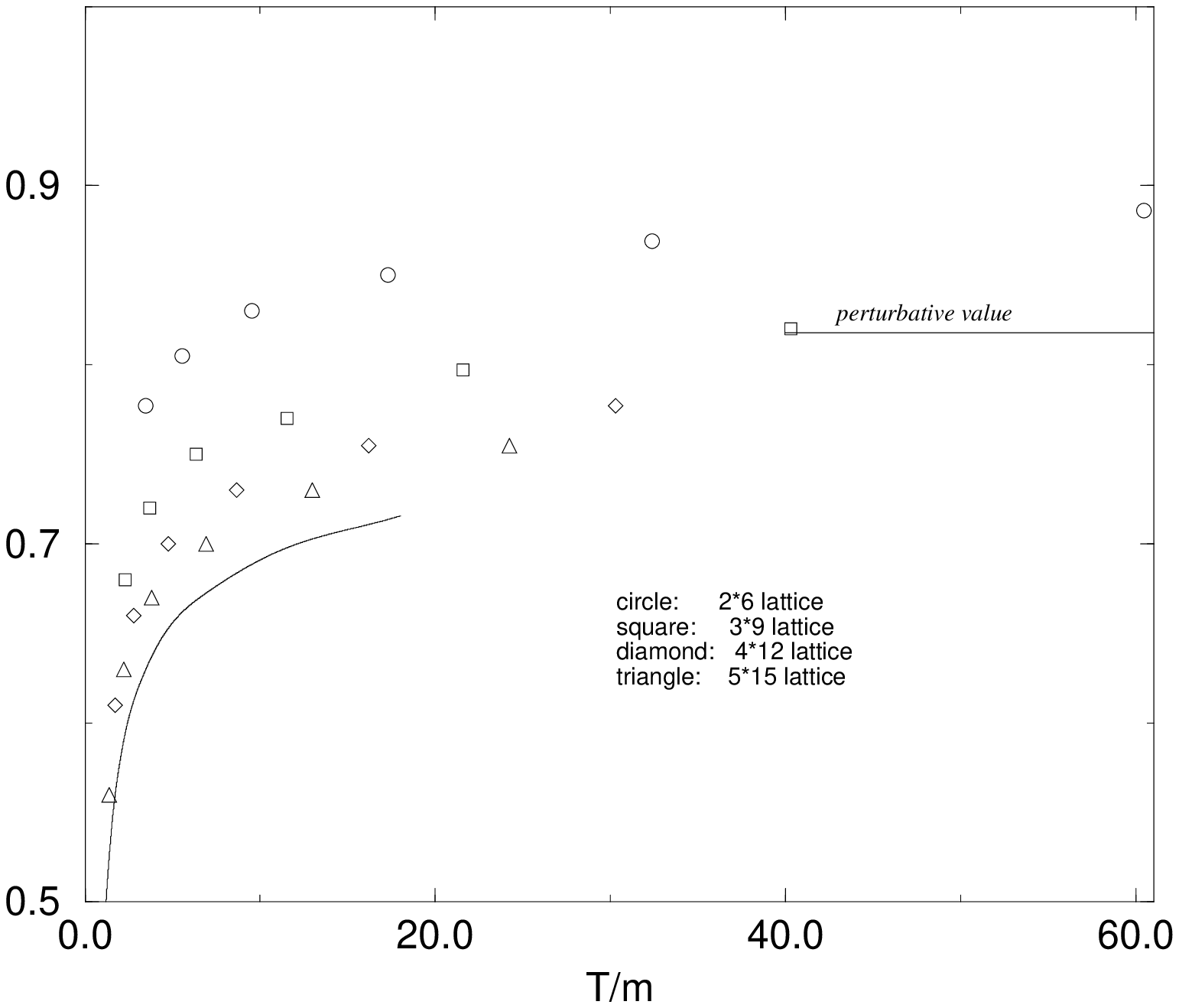}
\caption{The discreet points give the free energy density
$-{\mathbf f}/T^{2}$ for different lattice resolutions with
Wilson action, whereas the smooth curve is a spline 
interpolation of the data resulting from a $3 \times 9$
lattice with FP action. The error of
the free energy density is smaller than the 
size of the symbols. The horizontal line
shows the perturbative value for $\zeta=3$ in the high
temperature limit.}
\label{fig:Stand}
\vspace*{6ex}
\end{center}
\end{figure}

\begin{figure}[h]
\begin{center}
\leavevmode
\epsfxsize=0.8\textwidth
\epsfbox{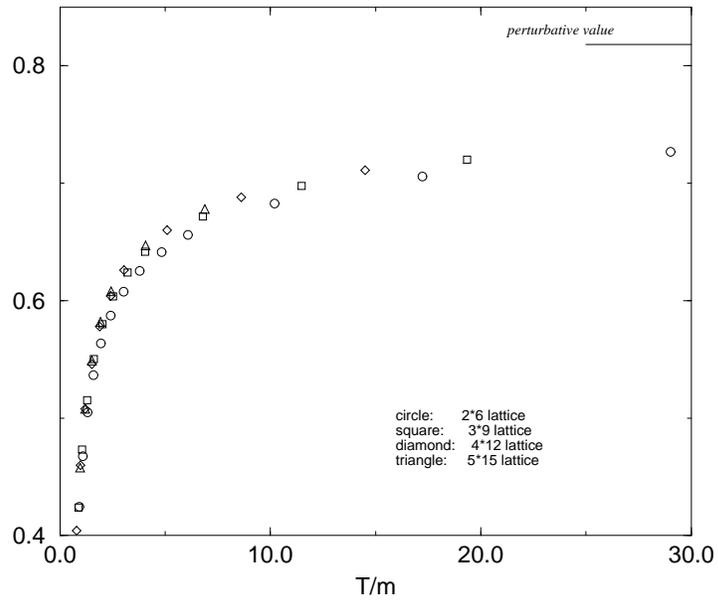}
\caption{The free energy density for different lattice resolutions
using FP action. The error of these values is smaller than
the size of the symbols.}
\label{fig:Fix}
\vspace*{6ex}
\end{center}
\end{figure}

\begin{figure}[h]
\begin{center}
\leavevmode
\epsfxsize=0.8\textwidth
\epsfbox{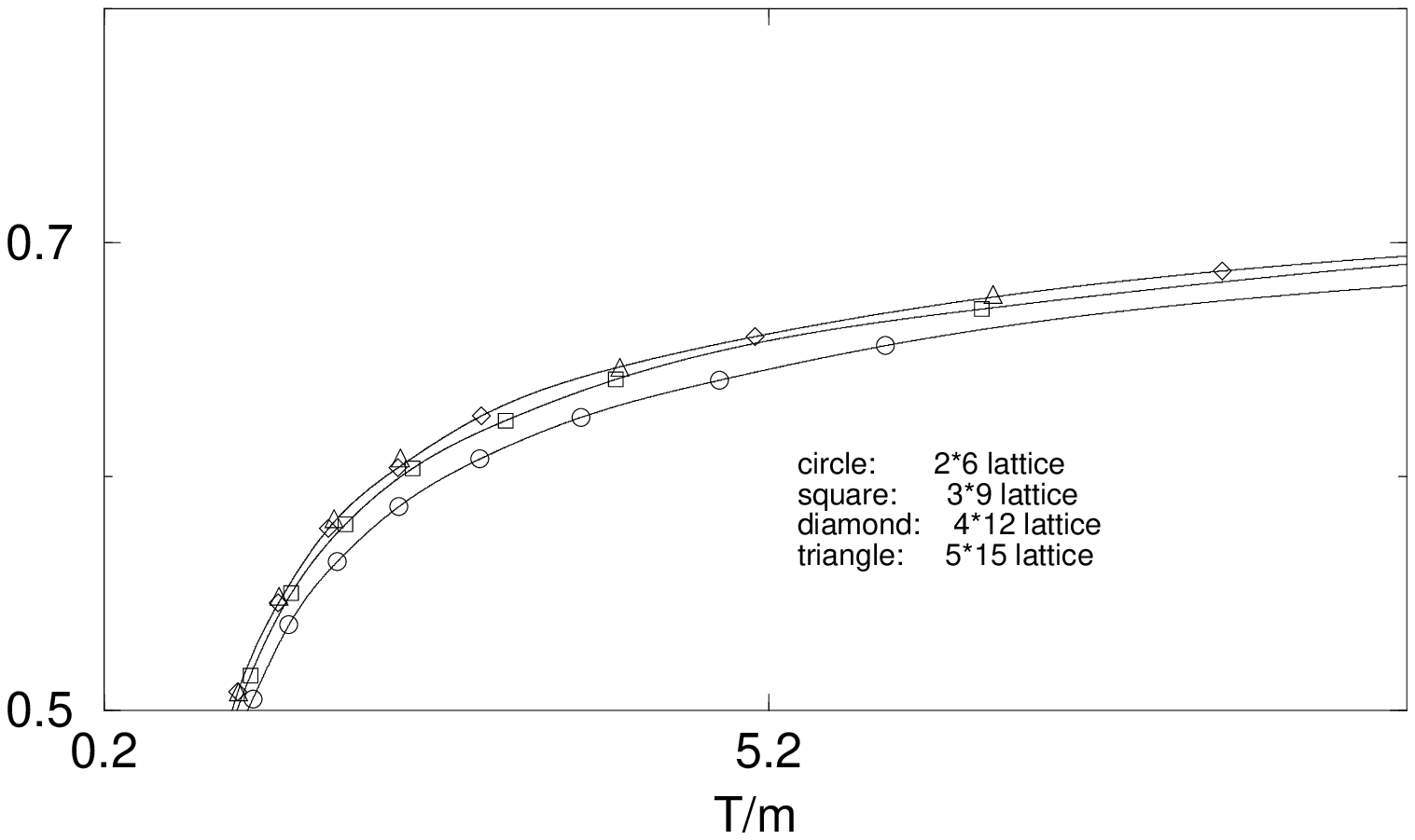}
\caption{The smooth curves show a spline interpolation of the
data for $2 \times 6$, $3\times9$ and $4\times12$ lattices in 
the FP case (fig. \ref{fig:Fix}). Within 
 statistical error
the values for the $5\times15$ lattice lie on the curve
of the $4\times12$ lattice.}
\label{fig:last}
\vspace*{6ex}
\end{center}
\end{figure}

 In our simulations we used a
cluster algorithm \cite{cluster}, which almost completely
 eliminates the critical
slowing--down,  which was
especially important for the measurement of the mass gap. For the
measurement of the action we  performed
$8\times 10^{5}$ sweeps per $\beta$--value and estimated the
error by the method of bunching. The size of the error is of the order
of the circles in figure \ref{fig:deltas}, where one finds the signal
of the integrand in eq.
(\ref{eq:thermo}). 
 The smooth curve  is  a spline fit interpolation,
which
we used for the integration, where  $\beta_{0}$ had to be chosen so
small that the signal for that value disappeares.
 To estimate the error of this integration
we compared the value resulting from the spline curve with the
value one gets, joining the single points in fig. \ref{fig:deltas}
by straight lines. This error is not bigger than half the size of
the symbols in figures \ref{fig:Stand}, \ref{fig:Fix} and
\ref{fig:last}. To compare the cut--off effects we measured the
action for Wilson and FP parametrization.

The above procedure  gives just the free energy density as a
function of $\beta$.
 To extract physical results we must relate
every $\beta$--value to a temperature. This can be done by
measuring another  physical quantity, the mass
gap. In a lattice simulation one can only get the dimensionless
quantity $m\cdot a$, where the lattice spacing $a$ is a function
of $\beta$. Using the relation $T=1/N_{\mathrm t}a$ we find:
\begin{equation} \label{eq:mass}
\frac{T}{m} = \frac{1}{maN_{\mathrm t}}
\end{equation}
 where $N_{\mathrm t}$ is the number of points in the time direction of
the lattice used for the measurement of $\langle
{\mathcal A}_{\mathrm t} \rangle$.
 We then extracted
 the quantity $m \cdot a(\beta)$ for some $\beta$--values by taking
a square lattice which was about seven times larger than the
correlation length to avoid finite size effects. Inserting these values
in eq. (\ref{eq:mass}) we got a finite number of points $T/m\;(\beta)$.
To our  data we added further results of the mass gap in the 
FP case from ref. \cite{Ruedi}, while for the Wilson action
we used the results from \cite{Improved}.

We present first (fig. \ref{fig:Stand}) a comparison between 
Wilson action and FP action. Fig. \ref{fig:Fix} gives
the detailed results obtained with the FP action and
the last figure (fig. \ref{fig:last}) shows a spline interpolation
of the FP data.

\section{Summary}
The most important result of this work is the great reduction of
cut--off effects with the FP action compared to other
lattice actions. In the perturbative regime the cut--off effects
disappear exponentially fast and are nearly lost on a lattice
as small as $N_{\mathrm t}=4$. Concerning the cut--off effects,
 our results of MC simulations show within the range of
statistical error
a similar  behaviour in the
low--temperature region. Although we are working with a relatively
small statistical error we are not able to detect any power--law
in the decrease of the cut--off effects.

\begin{ack}
My special thanks is directed to P.~Hasenfratz for his continuous
help during this work. I am indebted to P.~Hasenfratz and 
F.~Niedermayer for allowing me to use their cluster algorithm. 
 I would also like to thank F.~Niedermayer, 
M.~Blatter, R.~Burkhalter, A.~Papa and Ch.~Stulz for useful
discussions and the entire institute in Bern for their kind
hospitality. I should not forget D.~Wyler who stimulated my interest
in lattice field theory.
\end{ack}


\begin{thebibliography}{99}

\bibitem{Kogut}
K.~Wilson and J.~Kogut, Phys.\ Rep.\ C12 (1974) 75 \\
K.~Wilson, Rev.\ Mod.\ Phys.\ 47 (1975) 773, 55 (1983) 583

\bibitem{ Wilson}
K.~Wilson, {\textit {in}} Recent developments of gauge theories, ed.
G.'t~Hooft et al. (Plenum, New York, 1980)


\bibitem{Hase1}
P.~Hasenfratz, F.~Niedermayer, Nucl.\ Phys.\ B414 (1994) 785

\bibitem{cut}
A.~Papa, Nucl.\ Phys.\ B 478 (1996) 335

\bibitem{Boyd}
G. Boyd et al., Phys.\ Rev.\ Lett.\ 75 (1995) 4169 
and references therein

\bibitem{Ruedi}
M.~Blatter, R.~Burkhalter, P.~Hasenfratz, F.~Niedermayer,
Phys.\ Rev.\ D53 (1996) 923

\bibitem{Leutwyler}
P.~Hasenfratz, Phys.\ Lett.\ 114B (1982) 251 \\
P.~Hasenfratz, H.~Leutwyler, Nucl.\ Phys.\ B343 (1990) 241

\bibitem{cluster}
R.~H.~Swendsen, J.~S.~Wang, Phys.\ Rev.\ Lett.\ 58 (1987) 86 \\
U.~Wolff, Phys.\ Rev.\ Lett.\ 62 (1989) 361


\bibitem{Improved}
U.~Wolff, Nucl.\ Phys.\ B334 (1990) 581--610


\end{thebibliography}
\end{document}